\documentclass[]{emulateapj}
\usepackage{graphicx}
\usepackage{latexsym}
\usepackage{natbib}

\begin{document}

\title{Discovery of a 205.89~Hz accreting--millisecond X-ray pulsar in
  the globular cluster NGC~6440}

\author{D. Altamirano\altaffilmark{1}
A. Patruno\altaffilmark{1},
C.~O. Heinke\altaffilmark{2},
C. Markwardt\altaffilmark{3,4},
T.~E. Strohmayer\altaffilmark{3},\\
M. Linares\altaffilmark{1},   
R. Wijnands\altaffilmark{1},  
M. van der Klis\altaffilmark{1},
J.~H. Swank\altaffilmark{3}}

\altaffiltext{1}{Email: d.altamirano@uva.nl ; Astronomical Institute,
  ``Anton Pannekoek'', University of Amsterdam, Science Park 904,
  1098XH, Amsterdam, The Netherlands.}

\altaffiltext{2}{University of Alberta, Department of Physics, Room
  238 CEB, 11322-89 Avenue, Edmonton, AB T6G 2G7, Canada.}

\altaffiltext{3}{Laboratory for High-Energy Astrophysics, NASA Goddard
  Space Flight Center, Greenbelt, MD 20771, U.S.A.}

\altaffiltext{4}{Department of Astronomy, University of Maryland,
  College Park, MD 20742., U.S.A. }

\begin{abstract}

  We report the discovery of the second accreting millisecond X-ray
  pulsar (AMXP) in the globular cluster NGC~6440. Pulsations with a
  frequency of 205.89~Hz were detected with the Rossi X-Ray Timing
  Explorer on August 30th, October 1st and October 28th, 2009, during
  the decays of $\lesssim4$ day outbursts of a newly X-ray transient
  source in NGC~6440.
  By studying the Doppler shift of the pulsation frequency, we find
  that the system is an ultra-compact binary with an orbital period of
  57.3 minutes and a projected semi-major axis of 6.22
  light-milliseconds.
  Based on the mass function, we estimate a lower limit to the mass of
  the companion to be 0.0067 M$_{\odot}$ (assuming a 1.4 M$_{\odot}$
  neutron star).
  This new pulsar shows the shortest outburst recurrence time among
  AMXPs ($\sim1$ month). If this behavior does not cease, this AMXP
  has the potential to be one of the best sources in which to study
  how the binary system and the neutron star spin evolve.
  Furthermore, the characteristics of this new source indicate that
  there might exist a population of AMXPs undergoing weak outbursts
  which are undetected by current all-sky X-ray monitors.
  NGC~6440 is the only globular cluster to host two known AMXPs, while
  no AMXPs have been detected in any other globular cluster.

\end{abstract}


\maketitle

\section{Introduction}
\label{sec:intro}

Accreting millisecond X-ray pulsars (AMXPs) are rapidly spinning
neutron stars accreting matter from a low-mass stellar
companion. These systems are thought to be the progenitors of the
millisecond radio pulsars \citep{Alpar82,Backer82}. Therefore, 
their study allows us to better understand how neutron stars in
low-mass X-ray binaries (LMXB) evolve into radio millisecond
pulsars. Furthermore, AMXPs are perfect laboratories to study the
interaction between the neutron star magnetosphere and the accretion
disk \citep[see, e.g.,][and references therein]{Poutanen06}.

A total of ten AMXPs among the $\sim100$ neutron star LMXBs were known
by August 2009.
Seven of these AMXPs showed outbursts lasting a few days to months,
and recurrence times of at least ~2 years.
In all seven cases, the pulsations were detected persistently
during the whole outburst.
The remaining three AMXPs showed pulsations only intermittently
\citep{Galloway07a,Gavriil07,Casella08,Altamirano08b,Patruno09},
bridging the gap between the small number of AMXPs and the large group
of non-pulsating LMXBs \citep{Altamirano08b}. The reason why only a
small amount of neutron star LMXBs show pulsations is still under
debate.
One of the proposed scenarios for the non-pulsating systems is that
the neutron-star magnetic field could be temporarily buried by the
accreted matter \citep{Cumming01}.  If the time-averaged mass
accretion rate of the accretors is relatively high, the accreted
matter can bury the field throughout the life of the X-ray binary so
the neutron star does not pulsate. However, if the average accretion
rate is low enough that the magnetic field can diffuse through the
accreted matter faster than it is buried, then these systems will
probably exhibit pulsations. 
Another of the proposed scenarios that explains why the majority of
neutron stars in LMXBs do not pulsate is that of \citet{Lamb09a}; if
neutron stars undergo long periods of accretion (at high rates), then
their magnetic poles are naturally forced to
align to their spin axes as they spin up.  When the accretion rate decreases, the
magnetic poles can move away from the rotation axis (due to magnetic
dipole and other braking torques which cause neutron stars to spin
down) and oscillations powered by accretion should become visible.
Both \citet{Cumming01} and \citet{Lamb09a}  present clear
predictions: we should find pulsations in many (if not all) of the
systems which show low time-averaged mass accretion rates.
Very recently, a Chandra X-ray observation serendipitously discovered
a new X-ray transient in the globular cluster NGC~6440
\citep[identified as NGC~6440 X-2]{Heinke09a, Heinke09b,
  Heinke09c}. RXTE and Swift follow-up observations not only showed
that this source has a very low time--average accretion rate
\citep[$\lesssim2 \cdot 10^{-12}$ M$_{\odot}$ year$^{-1}$, see
][]{Heinke09d} but, as predicted by the models discussed above, also
millisecond X-ray pulsations \citep{Altamirano09}.
One of the most interesting properties of this new AMXP is its
very short recurrence time \citep[$\sim1$ month, see
also][]{Heinke09d}. If its outbursts continue, it will be possible to study the
evolution of spin and orbital parameters on timescales never before probed.

In this Letter we report the discovery of X-ray pulsations in NGC
6440 X-2.
In a companion work \citep{Heinke09d}, we report on a detailed
description of the evolution of the source's outbursts based on our
multiwavelength campaign on this source.
Our results demonstrate that there may exist a population of AMXPs
undergoing low-luminosity outbursts which are undetected by current
all-sky X-ray monitors.

\section{LMXBs in the globular cluster NGC~6440} 

NGC~6440 is a globular cluster at $8.5\pm0.4$~kpc \citep{Ortolani94}.
Using Chandra images, \citet{Pooley02} studied the population of the
low-luminosity X-ray sources in NGC~6440. These authors found 24 X-ray
sources above a limiting luminosity of $\sim2 \cdot 10^{31}$ ergs
s$^{-1}$ (0.5--2.5 keV) inside the half-mass radius of the cluster
(all of which lie within $\sim$2 core radii of the cluster center); there
is strong evidence that 8 of these systems are LMXBs in quiescence
\citep{Heinke03}. \citet{Pooley02} also found excess emission in and
around the core of NGC~6440, suggesting unresolved point
sources.

Bright X-ray outbursts from a LMXB were observed in 1971, 1998, 2001
and 2005 \citep{Markert75,Zand99,Verbunt00,Zand01,Markwardt05}.
From X-ray and optical observations, \citet{Zand01} concluded that the
1998 and 2001 outbursts were from the same object, which they
designated SAX~J1748.9--2021.  \citet{Altamirano08b} detected
intermittent X-ray pulsations at $\sim442$~Hz in the 2001 and 2005
outbursts \citep[see also ][]{Gavriil07}.  Whether the 1971 outburst
was from SAX~J1748.9--2021 is uncertain.

Between MJD 54875 and 55075 at least 5 short-duration ($\sim$days)
outbursts have been detected in NGC~6440.  As discussed in
\citet{Heinke09d}, Chandra, RXTE and Swift data indicates that all
outbursts were from the same source, denoted NGC~6440 X-2.

\begin{figure}[t] 
\center
\resizebox{1\columnwidth}{!}{\rotatebox{-90}{\includegraphics{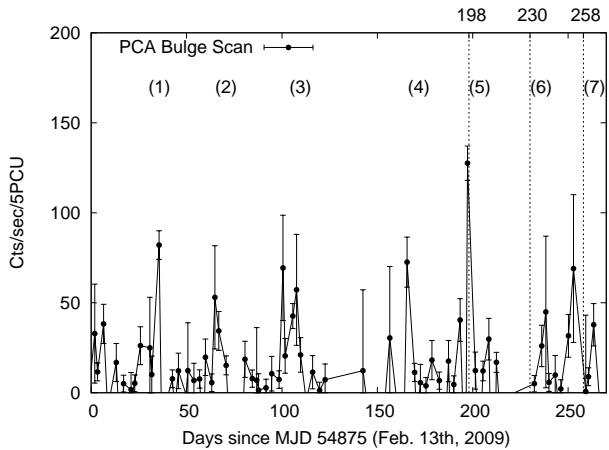}}}
\caption{Intensity (2.0--10.0~keV cts/sec/5PCU) vs. time as measured
  by the PCA Bulge Scan monitoring \citep{Swank01}. The dashed lines
  mark the times when we detected X-ray pulsations (August 30th,
  October 1st and October 28th, 2009); note that the bulge scans did
  not sample the October 1st outburst. See \citet{Heinke09d} for more
  details.}
\label{fig:lc}
\end{figure}

\section{Observations, data analysis and results}
\label{sec:dataanalysis}

We used data from the RXTE Proportional Counter Array \citep[PCA, for
instrument information see][]{Jahoda06}. Between August 1998 and
November 4th, 2009, there were 49 pointed observations of the globular
cluster NGC~6440 each covering 1 to 5 consecutive 90-min satellite
orbits. Usually, an orbit contains between 1 and 5 ksec of useful data
separated by 1--4 ksec data gaps due to Earth occultations and South
Atlantic Anomaly passages.
The first 27 observations sample 3 outbursts of the Intermittent AMXP
SAX J1748.9--2021 \citep{Altamirano08b}. The remaining 22 were
triggered based on results of the PCA Bulge Scan program or ToO Swift
observations, and sample the 5 outbursts of NGC~6440 X-2 \citep[see
also][]{Heinke09d}.

We performed a timing analysis using the high-time resolution data
collected in the Event (E\_125us\_64M\_0\_1s), Good Xenon and
Single-Bit modes.  Fourier power spectra were constructed for each 
observation, using data segments of 128, 256 and 512 seconds and with
a Nyquist frequency of 4096~Hz.  No background or dead-time
corrections were made prior to the calculation of the power spectra,
but all reported rms amplitudes are background corrected; deadtime
corrections are negligible.

\begin{table}
\caption{Timing parameters for NGC~6440}

\scriptsize
\begin{tabular}{lc}
\hline
\hline
Parameter & Value \\
\hline
Orbital period, P$_{orb}$(seconds) \dotfill                           & 3438(33)  \\
Projected semi major axis, $a_x sin i$ (lt-ms)\dotfill   & 6.22(7)  \\
Time of ascending node, $T_{asc}$ (MJD/TDB) \dotfill          & 55073.0344(6) \\
Eccentricity, e (95\% c.l.)\dotfill                                         & $<0.07$ \\
Spin frequency $\nu_0$ (Hz) \dotfill                             & 205.89215(2)  \\
Pulsar mass function, $f_x$ ($\times 10^{-7} M_{\sun}$)\dotfill  & $1.6(1)$ \\
Minimum companion mass$^1$, $M_c$ ($M_{\sun}$)\dotfill               & $0.0067$ \\
\hline
\end{tabular}\\
All errors are at $\Delta\chi^2=1$. \\
$^1$: The minimum companion mass is estimated assuming a neutron star mass of 1.4 M$_{\odot}$.
\label{table:data}
\end{table}

%
%
%
%
%
%

\begin{figure}[!hbtp] 
\resizebox{1\columnwidth}{!}{\rotatebox{0}{\includegraphics{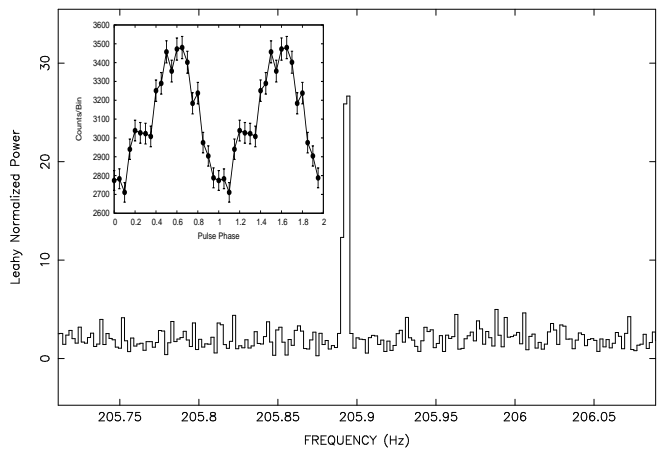}}}
\caption{Leahy normalized \citep{Leahy83} power spectrum of August 30th
  observation ($\sim3$~ksec of data).
  The signal is distributed on at least three independent frequency
  bins of size 1/512 Hz. \textit{Inset}: Pulse profile in the 2--14.8
  keV range (two cycles are plotted for clarity). The fractional rms
  amplitudes of the fundamental and first overtone are $7.5\pm0.6$\%
  and $2.3\pm0.6$\% rms, respectively. Count rates (in counts/bin) are
  corrected for background.  }
\label{fig:pds}
\end{figure}

\subsection{Light curves and power spectra}

In Figure~\ref{fig:lc} we plot the intensity (2.0--10.0~keV
cts/sec/PCA) of the globular cluster NGC~6440 as measured with PCA
Bulge Scans \citep{Swank01}. The data suggest that between MJD 54875
and 55075 the globular cluster has brightened for short periods
($\sim$days) on $\sim5$ occasions \citep[see
also][]{Heinke09d}. However, we note that some of these points might
be spurious given that the intensity measured is systematically
uncertain due to the contributions from diffuse galactic emission and
other nearby sources in the PCA field of view.
In none of the pointed nor the Bulge Scan observations did we detect
thermonuclear X-ray bursts.

As can be seen in Figure~\ref{fig:lc}, on August 30th, 2009 (MJD
55072) the PCA bulge scan detected the highest intensity in the
direction of NGC~6440 over a period of more than 250 days.
A ToO RXTE observation (ObsID: 94044-04-02-01) found NGC~6440 at a
luminosity of $L_{2-10 {\rm keV}}\sim1.7 \cdot 10^{36}$ ergs s$^{-1}$
\citep[using a distance of 8.5 kpc, see][]{Ortolani94}. The energy
spectrum is well fitted with an absorbed power law with index of
$1.83\pm0.3$ ($\chi^2/dof=0.84$ for 47 degrees of freedom; the
interstellar absorption $N_h$ was fixed to a value of $5.9 \cdot
10^{21}$ cm$^{-2}$, consistent with the cluster value and the
absorption measured by Chandra, see \citealt{Heinke09d}).
The power spectrum of this observation shows two features with
characteristic frequencies \citep{Belloni02} of $7.4\pm0.8$ and
$1.13\pm0.06$ Hz, quality factor Q of $1.3\pm0.6$ and $0.71\pm0.07$,
and rms amplitudes of $28\pm3$\% and $42\pm2$\%, respectively. We
searched for kHz QPOs, but found none. The average
(background-subtracted) count rate during this observation was
$\sim22$ cts/sec and only one proportional counter unit (PCU) was
active (2--60 keV).
Additional RXTE observations were performed on the 1st and 2nd of
September; fluxes were consistent with background emission. These
measurements are consistent with the fluxes measured by Swift between
the 1st and the 4th of September \citep{Heinke09d}. If we take into
account that a PCA bulge scan observation did not detect significant
flux in the direction of NGC~6440 on August 28th, we can estimate an
outburst duration of no more than $\sim3$ days (above $\sim10^{35}$
erg s$^{-1}$).

On July 30th, October 1st and October 28th, 2009, RXTE sampled three
other outbursts of NGC~6440 \citep{Heinke09d} for
$\sim1.5-2.5$~ksec. We did not find any significant QPOs nor broad
band noise in their power spectra, probably due to poor statistics
(note that the total count rates measured are dominated by background
counts).  These observations ranged from $L_{2-10 {\rm
    keV}}=1\times10^{35}$ to $5.2\times10^{35}$ ergs s$^{-1}$, thus
providing poorer statistics than on August 30th.
We found no evidence for dips or eclipses in any of the RXTE pointed
observations of NGC~6440 X-2.

\begin{figure}[!hbtp] 
\resizebox{1\columnwidth}{!}{\rotatebox{0}{\includegraphics{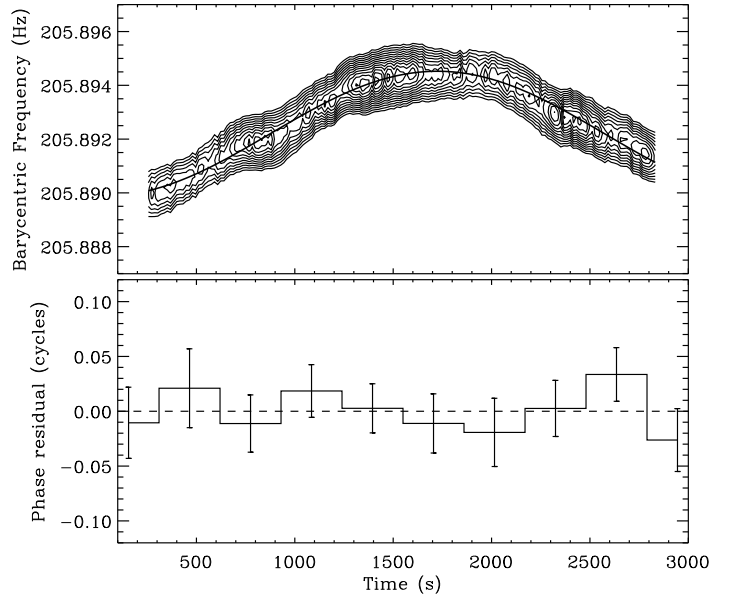}}}
\caption{\textit{Upper panel:} Dynamical power spectrum of August 30th
  observation with a 512 sec sliding window. The frequency is
  modulated as expected from the Doppler shifts due to the orbital motion of a
  pulsar in a binary system.  The data sample a little less than a
  full orbital cycle. The filled line represents our best orbital
  solution. \textit{Lower panel:} Phase residuals as estimated from
  300 sec data segments. All errors are at $\Delta\chi^2=1$. }
\label{fig:dyna}
\end{figure}

\subsection{Pulsations}

Adopting a source position $\alpha=17^h 48^m 52^s.75$, $\delta = -20^o
21^{'} 24^{''}.0$ \citep[from Chandra images, see][]{Heinke09d}, we
converted the photon arrival times to the Solar System barycenter with
the FTOOL faxbary, which uses the JPL DE-405 ephemeris along with the
spacecraft ephemeris and fine clock corrections to provide an absolute
timing accuracy of ~3.4 $\rm\,\mu s$ \citep{Jahoda06}.

We found a strong signal at a frequency of $\sim205.89$~Hz in the
August 30th RXTE observation. In Figure~\ref{fig:pds} we show the
Leahy normalized power spectrum.  We calculated a dynamical power
spectrum with a 512 sec sliding window and found that the pulse
frequency is modulated in a way that is typical from Doppler shifts
due to the orbital motion of a pulsar in a binary system (see
Figure~\ref{fig:dyna}); the data sample a little less than a full
orbital cycle.

To improve the signal to noise of the pulsations, we first modeled the
pulse drifts with a sinusoid (assuming zero eccentricity).
We found that the frequency drift is consistent with a system with an
orbital period of 57 minutes and a projected semi-major axis of 6.2
lt-ms.  With this provisional solution, we folded the 3 ks of
lightcurve into 10 pulse profiles of $\sim 300$ s each and fitted them
with a constant plus two sinusoids representing the pulse frequency and
its first overtone. 
We then phase-connected the pulse phases by fitting a constant pulse
frequency plus a circular Keplerian orbital model.  The procedure is
described in detail in \citet{Patruno09}.
In Table~\ref{table:data} we report the best fit.
In the inner panel of Figure~\ref{fig:pds} we show the folded pulse
profile in the 2--14.8 keV range (i.e. channels 0--35) after
correcting for the effects of the orbital modulation. The fractional
rms amplitudes of the fundamental and first overtone are $7.5\pm0.6$
and $2.3\pm0.6$\%, respectively (2.0--14.8 keV).

We applied the above technique to three energy bands: 2--5.7 keV,
6.1--14.8 keV and 15.2--60 keV.
(We note that we could only choose these bands, as the data were split
in two Single-bit modes -- SB\_125us\_0\_13\_1s \&
SB\_125us\_14\_35\_1s -- and one Event mode -- E\_62us\_32M\_36\_1s.)
The pulsed fractional rms amplitudes of the fundamental were
$(7.5\pm1.0)$\% and $(8.2\pm1.0)$\% in the 2-5.7 keV and 6.1-14.8
energy band.
The first overtone was significantly detected in the latter band, with
an amplitude of $3.3\pm0.8$\% rms.
Due to the low quality of the data at high energies, only upper limits
of 21\% rms (at 95\% confidence level) can be put on the fundamental
in the 15.2-60 keV band.
(NGC~6440 X-2 was more than 100 times brighter than the other 24 X-ray
sources in NGC 6440 \citep{Pooley02}, so their contributions can be
ignored. However, Galactic Ridge emission does provide additional
background at this location, not included in standard RXTE background
estimates, and therefore the rms amplitudes we quote above might be
slight underestimates).

We also searched for pulsations in the observations on July 30th, October
1st and October 28th, 2009, by correcting for the orbital motion of
the system and folding the data around the best spin period
determination.  No pulsations were found on July 30th, with 95\%
c.l. upper limits of 8\% rms on the pulse fractional amplitude. On
October 1st and 28th we found $3.4\sigma$ and $8.9\sigma$ detections,
respectively, of pulsations with characteristics consistent with those
we found on August 30th.


RXTE observed NGC~6440 during the three outbursts of the AMXP
SAX~J1748.9--2021 \citep[see, e.g.,][]{Altamirano08b,
  Patruno09}. Although improbable, it is possible that NGC~6440 X-2
was also active during these periods. To further investigate this, we
searched all previous RXTE observations of NGC~6440 for the presence
of pulsations at $\sim205-206$~Hz. We barycentered all data in the
same manner as above and divided the data in 3 groups, corresponding
to the 1998, 2001 and 2005 outbursts. For each group we performed
Fourier transforms of all the data and calculated a single averaged
power spectra. We found no significant pulsations in the
$\sim205-206$~Hz range.
We also searched for evidence of intermittent pulsations such as those
found in other AMXPs \citep{Galloway07a, Gavriil07, Altamirano08b,
  Casella08}, but found none. 
In the above searches, we did not correct for the orbital motion of
the binary; although these corrections would improve our sensitivity,
such analysis is beyond the scope of this Letter.

Finally, we note that it was not possible to search for pulsations in
the Swift observations. These data-sets were obtained in
photon-counting mode with a time resolution of 2.5 seconds.

\section{Discussion}\label{sec:discussion}

We have discovered the second accreting millisecond X-ray pulsar in
the globular cluster NGC~6440. Pulsations were detected in three RXTE
pointed observations performed during the decay phase of three outbursts, each of which lasted less than $\sim3$ days \citep{Heinke09d}.
The frequency drifts we observe are consistent with an orbital period
of 57.3 minutes and a semi-major axis ($a_x$ sin$i$) of 6.22
light-msec, which shows that this new AMXP is in an ultra-compact
binary system.

The characteristics of this new pulsar are similar to those of other
AMXPs, particularly SWIFT J1756.9--2508 \citep{Krimm07,Patruno10},
which showed pulsations at a frequency of 182~Hz, an orbital period of
54.7 minutes, a projected semi-major axis of 5.94 lt-ms and a minimum
companion mass of 0.0067 M$_{\odot}$. 
As pointed out by \citealt{Krimm07}, such a low mass for a companion
in this type of system is probably inconsistent with brown dwarf
models while white dwarf models suggest that the companion is probably
a He-dominated donor.

The reason why only twelve\footnote{See also the recent discovery of a
  245~Hz AMXP by \citet{Markwardt09}.} NS LMXBs exhibit coherent
pulsations out of about 100 systems is still unclear.  There is a
strong tendency for the detected AMXPs to have rather low
time-averaged accretion rates \citep[over many outbursts, see, e.g.,
][]{Chakrabarty05,Galloway06a,Wijnands08}. The time-averaged accretion
rate of NGC~6440 X-2 is at most similar to that of the AMXP
SAX~J1808.4--3658, but could easily be much lower \citep[see
discussion by ][]{Heinke09d}. This is consistent with the general
tendency and furthermore, with the theoretical predictions
\citep{Cumming01, Lamb09a} that a significant fraction (if not all) of
the NS LMXB systems with low time-averaged accretion rates harbor
accreting millisecond X-ray pulsars \citep{Wijnands08}.
The outburst durations ($\sim3$ days) of the newly discovered AMXP are
short compared with those of the other AMXPs, although not uniquely
so. For example, short-duration outbursts of the AMXP XTE J1751--305
have been detected in 2005, 2007 and 2009 \citep{Grebenev05,Swank05,
  Markwardt07b, Markwardt07c,Linares07a, Markwardt09a}.
Our discovery of NGC~6440 X-2 indicates that there might exist a
population of AMXPs undergoing weak outbursts. 
Unfortunately, short low-luminosity outbursts cannot be detected with
the current all-sky X-ray monitors (e.g., ASM aboard RXTE). More
sensitive instruments like the ``Monitor of All-sky X-ray Image''
\citep[MAXI, which should detect X-ray -- 2--30 keV -- sources of
about 20 mCrab from 90 minutes observations, and sources of about 4.5
mCrab after a day observation, see][]{Matsuoka09} will probably allow
us to discover many more potential low-luminosity AMXPs.

As discussed in \citet{Heinke09d}, NGC~6440 X-2 has been in outburst
at least five times in the last 250 days. If true, this is the AMXP
with the shortest recurrence time. If NGC~6440 X-2 continues
undergoing this type of short outbursts and RXTE is able to observe
them, then NGC~6440 X-2 has the potential to be the best millisecond
X-ray pulsar in which to study the evolution of the binary as well as
that of the neutron star spin (by means of coherent timing analysis).

The globular cluster NGC~6440 is known for having at least 24 faint
X-ray sources to a limiting luminosity of $\sim2 \cdot 10^{31}$ ergs
s$^{-1}$ \citep[0.5--2.5 keV, see ][]{Pooley02}. NGC~6440 is also
known for hosting at least 6 millisecond radio pulsars, of which 3
are in binary systems \citep{Freire08}. None of the known radio
pulsars match the position (nor the characteristics) of NGC~6440 X-2.
Although millisecond radio pulsars in binary systems have been found
in other globular clusters \citep[see, e.g.,][and references
therein]{Ransom05, Freire08}, it is an intriguing question why
NGC~6440 is the only globular cluster known today to host (two) AMXPs
(SAX~J1748.9--2021 -- \citealt{Gavriil07} and \citealt{Altamirano08b}
-- and NGC~6440 X-2 -- this paper), while no AMXPs have been detected
in any other globular cluster.
This might be only a selection effect, as current all-sky monitors
cannot detect short low-luminosity outbursts. Monitoring X-ray
observations of different globular clusters would be needed to further
investigate this.

\end{document}